\documentclass[sigconf]{acmart}
\AtBeginDocument{%
  }

\copyrightyear{2025}
\acmYear{2025}
\setcopyright{cc}
\setcctype{by-sa}
\acmConference[DIS '25]{Designing Interactive Systems Conference}{July 5--9, 2025}{Funchal, Portugal}
\acmBooktitle{Designing Interactive Systems Conference (DIS '25), July 5--9, 2025, Funchal, Portugal}\acmDOI{10.1145/3715336.3735845}
\acmISBN{979-8-4007-1485-6/2025/07}

\usepackage{makecell} 
\usepackage{stfloats}
\usepackage{multirow} 

\begin{document}

\title{NoRe: Augmenting Journaling Experience with Generative AI for Music Creation}

\author{Joonyoung Park}
\authornote{These authors contributed equally to this research.}
\orcid{0009-0008-5003-6673}
\affiliation{%
  \institution{Information Science and Culture Studies}
  \institution{Seoul National University}
  \country{Republic of Korea}
  \postcode{08826}
}
\email{jooony@snu.ac.kr}

\author{Hyewon Cho}
\authornotemark[1]
\orcid{0009-0002-0460-7259}
\affiliation{%
  \institution{Information Science and Culture Studies}
  \institution{Seoul National University}
  \country{Republic of Korea}
  \postcode{08826}
}
\email{cocohw@snu.ac.kr}

\author{Hyehyun Chu}
\authornotemark[1]
\orcid{0009-0006-0256-5277}
\affiliation{%
  \institution{School of Computing}
  \institution{KAIST}
  \country{Republic of Korea}
  \postcode{34141}
}
\email{hyenchu@kaist.ac.kr}

\author{Yeeun Lee}
\orcid{0009-0005-9281-3328}
\affiliation{%
  \institution{Information Science and Culture Studies}
  \institution{Seoul National University}
  \country{Republic of Korea}
  \postcode{08826}
}
\email{jigtm1107@snu.ac.kr}

\author{Hajin Lim}
\authornote{Corresponding author}
\orcid{0000-0002-4746-2144}
\affiliation{%
  \institution{Information Science and Culture Studies}
  \institution{Seoul National University}
  \country{Republic of Korea}
  \postcode{08826}
}
\email{hajin@snu.ac.kr}

\renewcommand{\shortauthors}{Park, Cho, Chu et al.}

\begin{abstract}
  Journaling has long been recognized for fostering emotional awareness and self-reflection, and recent advancements in generative AI offer new opportunities to create personalized music that can enhance these practices. In this study, we explore how AI-generated music can augment the journaling experience.
  Through a formative study, we examined journal writers’ writing patterns, purposes, emotional regulation strategies, and the design requirements for the system that augments journaling experience by journal-based AI-generated music.
  Based on these insights, we developed NoRe, a system that transforms journal entries into personalized music using generative AI. In a seven-day in-the-wild study (N=15), we investigated user engagement and perceived emotional effectiveness through system logs, surveys, and interviews. Our findings suggest that journal-based music generation could support emotional reflection and provide vivid reminiscence of daily experiences.
  Drawing from these findings, we discuss design implications for tailoring music to journal writers’ emotional states and preferences.
\end{abstract}




\maketitle

\begin{figure*}
  \includegraphics[width=\textwidth]{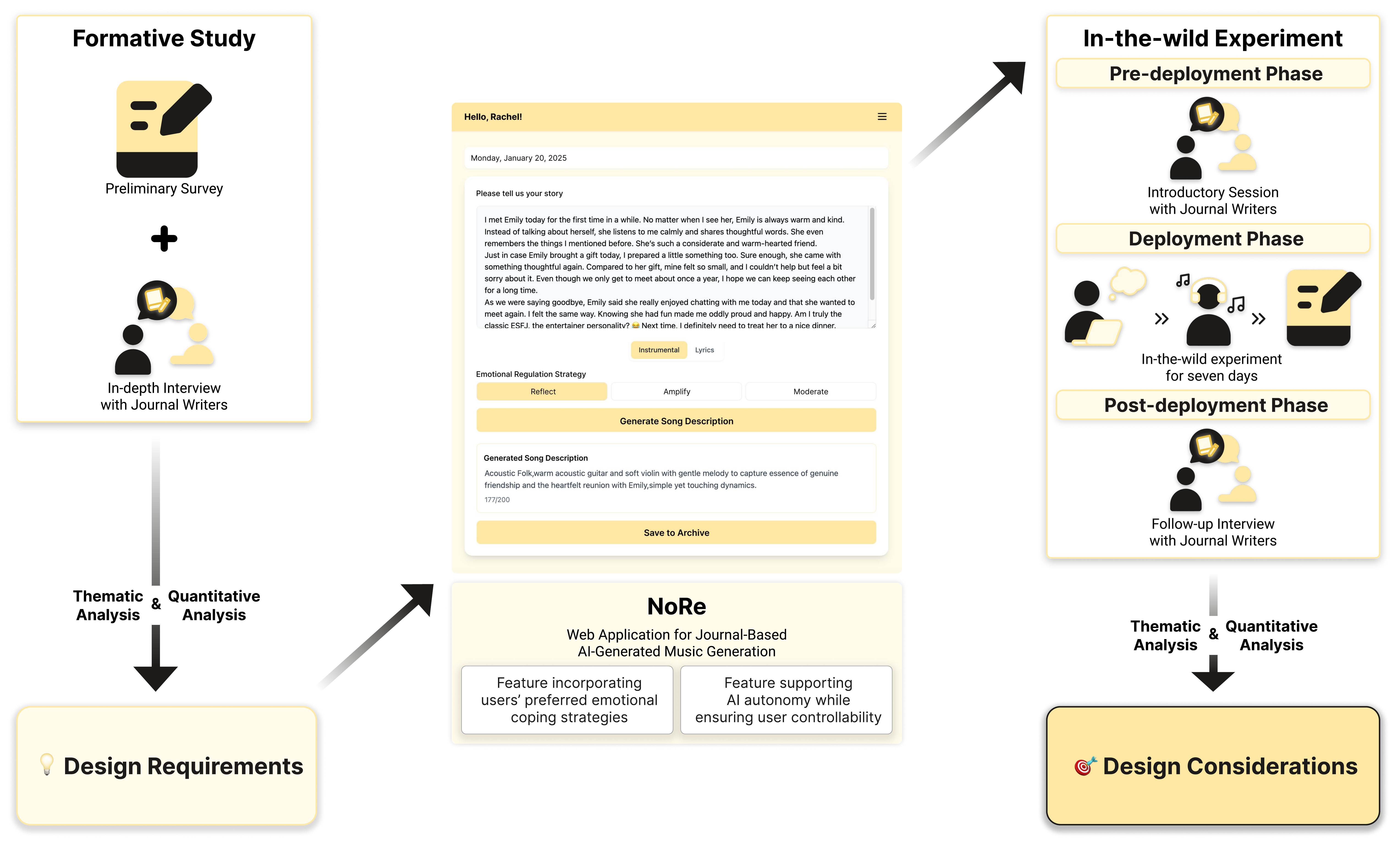}
  \caption{Research overview. This figure outlines the overall research procedure taken in this study. As a formative study, a preliminary survey and in-depth interviews were conducted with journal writers. Based on the thematic analysis and quantitative analysis, we derived design requirements that informed the development of NoRe, a web-based platform integrating journaling with AI-generated music. The platform incorporates features for emotional regulation strategies and user control over AI music generation. Following this, we conducted an in-the-wild evaluation over three phases: pre-deployment (introductory session), deployment (seven-day system usage and questionnaires), and post-deployment (follow-up interview). The study concluded with design considerations for future journal-based AI music generation systems.}
  \label{fig:teaser}
  \Description{The figure visualizes our research process in three main sections. The left section shows the formative study phase, consisting of preliminary surveys and in-depth interviews with journal writers, leading to design requirements through analysis. The center displays the NoRe web application interface, featuring a journal entry system with emotional regulation strategies and AI music generation controls. The right section illustrates the in-the-wild experiment phases: pre-deployment introductory sessions, seven-day system deployment, and post-deployment follow-up interviews, concluding with design considerations. Arrows connect these sections, showing the progression from initial research through system development to final evaluation.}
\end{figure*}

\section{Introduction}
Reflective writing has long been recognized as a valuable practice across domains such as education, healthcare, and the arts, supporting emotional awareness, personal growth, and professional development \cite{brandenburg_reflective_2017, ghaye_teaching_2010, korthagen_levels_2005, kolb1974toward, schon_reflective_1983}. Among these practices, journaling is a widely adopted method that provides structure for processing thoughts and emotions through writing \cite{patterson2011handbook, sohal_efficacy_2022}. By revisiting meaningful moments and reflecting on emotional responses, journaling enables individuals to recognize patterns, regulate emotions, and develop clarity and insight into their experiences \cite{lieberman2007putting, pennebaker1986confronting, alexander2016exploring}.



Like journaling, music can serve as a powerful medium for reflection and emotional regulation. Research shows that music helps individuals navigate and reshape their emotional experiences, especially when it resonates with their current emotional state \cite{thoma_emotion_2012}. Building on this, music has also been widely used in both clinical and everyday settings to promote psychological well-being. Studies highlight its physiological and therapeutic effects, positioning music as a valuable tool for introspection, stress reduction, and affective engagement \cite{fallon2020stress, de2022music, burrai2016randomized, akin2011internet}.


These reflective qualities of music have attracted interest in its integration with writing practices across various domains.
Prior research has shown that music can enhance writing performance \cite{ervin2002effect, donohoe_effect_1999} and influence the emotional tone and expressive depth of reflective writing \cite{kruse-weber_reflective_2023, hu_effects_2021}. 
These findings highlight music’s potential to support journaling by fostering deeper emotional engagement and self-reflection. 

However, most existing interventions rely on manually curated music selections, which limits their adaptability to individuals' unique emotional contexts \cite{baglione2021understanding}. This limitation reveals a gap between the recognized potential of music in reflective practices and its practical implementation in personalized contexts.

Recent advances in generative AI (GenAI) now enable the dynamic generation of music tailored to users’ emotional states and experiences \cite{hou2022ai}. This development opens up new possibilities for integrating AI-generated music into self-reflective practices by providing personalized, emotionally attuned soundscapes that support reflection and emotional regulation \cite{lee2024implementation}.

Despite these technical possibilities, the integration of AI-generated music into journaling remains largely unexplored. While reflective writing and music have each been extensively studied for their emotional and introspective benefits, little is known about how their combination, particularly when mediated by GenAI, might shape the journaling experience. This integration may offer unique opportunities for emotional processing: journaling provides a structured framework for articulating feelings through language, while AI-generated music grounded in journal entries could serve as a complementary modality for experiencing and engaging with those emotions through sound.


Building on this perspective, our study explores how a music generation AI-based system that transforms journal entries into music might enrich the reflective writing experience by fostering deeper emotional engagement. Through the design and deployment of this system, we aim to investigate how this integration might expand the expressive and emotional scope of journaling..

We structured our research in two phases.
In the first phase, we conducted a formative study with 15 regular journal writers to understand why and how they keep journals and to explore effective approaches for generating music from journal entries using GenAI.
Through in-depth interviews and evaluations of AI-generated musical pieces—produced using different prompting strategies based on their own journal entries—we examined participants’ journaling habits and goals, emotional regulation strategies, and preferences for how written experiences could be musically represented.


Our findings revealed that participants primarily used journaling for emotional processing and self-reflection.
Many expressed strong interest in transforming their emotional experiences into personalized music, emphasizing the value of music outputs that accurately captured the emotional nuances of their writing. Participants especially favored musical outputs generated through AI’s creative autonomy rather than rigid rule-based prompts, noting that such music felt more reflective of their entries. These insights informed a set of design requirements to support a range of reflective practices through the interplay between written expression and AI-generated music.

In the second phase of the study, building on these design requirements, we developed NoRe—a web-based platform that integrates journaling with AI-generated music. It includes features that support emotional regulation and give users control over the music generation process, enabling personalized and reflective musical interpretations of journal entries.

To explore NoRe's potential, we conducted an in-the-wild study with 15 participants over a seven-day period.
During the study, we collected and analyzed 50 journal entries alongside their corresponding AI-generated music, offering insights into how participants perceived the emotional alignment of the generated music, the strategies they employed to create music outputs, and how they incorporated the system into their journaling practice.

Our findings suggested that journal-based music generation could meaningfully support emotional reflection and engagement. Participants seamlessly integrated NoRe into their existing routines, demonstrating its compatibility with diverse journaling habits. Further, they viewed the system as a meaningful extension of their reflective process, highlighting the emotional relevance of the generated music in stimulating reminiscence, enhancing emotional regulation, and deepening self-understanding.
Based on these insights, we propose a set of design implications to inform future systems that integrate AI-generated music with self-reflection, contributing to broader discussions on the role of GenAI in self-care and reflective practices.


The contributions of this study are as follows:
\begin{itemize}
\item Empirical insights from a formative study with 15 journal writers, highlighting opportunities for integrating AI-generated music in journaling.
\item The design and implementation of NoRe, a web-based journaling system that integrates journaling with AI-generated music.
\item Findings from a seven-day field study, analyzing 50 journal entries and corresponding AI-generated music to understand how users incorporate AI-generated music into their journaling practices.
\item Design implications for the integration of AI-generated music with reflective writing.
\end{itemize}

\section{Related Work}
In this section, we cover the related work in three parts: (1) journaling for self-care, (2) music for mental health, and (3) potential of AI-generated music for self-reflection. 
 
\subsection{Journaling for Self-care}
Journaling is a widely practiced form of expressive writing that supports self-reflection by helping individuals articulate their thoughts, emotions, and experiences in a secure space \cite{riddell_healthy_2020}.
It has been associated with a range of psychological benefits, from enhancing self-awareness and emotional insight \cite{patterson2011handbook} to fostering metacognitive skills and higher-order thinking \cite{negretti2012metacognition}.
As a form of reflective practice, journaling enables individuals to revisit and make sense of personal experiences, contributing to emotional processing and personal growth \cite{schon_reflective_1983, kolb1974toward, ghaye_teaching_2010}.

Empirical studies have shown that journaling can reduce stress and depressive symptoms \cite{stice2007randomized, kerner2007integrating, smyth1998written}, mitigate rumination \cite{lodhi2018impact, boals2012use}, and support the development of emotional regulation strategies \cite{suhr2017maintaining, pennebaker1986confronting}.
It also serves as a tool for tracking emotional and behavioral patterns, making it valuable in both personal reflection and psychological research \cite{alexander2016exploring, travers2011unveiling, ullrich2002journaling}.
Beyond therapeutic contexts, journaling has been applied in professional settings to examine interpersonal dynamics and emotional well-being \cite{waddington2012using}.

Taken together, journaling serves a dual purpose: it supports ongoing psychological self-monitoring and provides a structured mechanism for self-reflection that fosters emotional awareness and regulation.
Through these combined effects, it has become a widely adopted strategy for cultivating personal insight and managing everyday emotional experiences.

Building on this potential, recent HCI and AI research has explored how journaling can be augmented through interactive systems. MindfulDiary \cite{kim_mindfuldiary_2024}, for example, uses a large language model (LLM) to guide psychiatric patients in recalling and articulating past emotions through dialogue, aiding mental health professionals in understanding their conditions.
Similarly, MindScape \cite{nepal_mindscape_2024} supports daily journaling and tracks users’ emotional states using LLMs.

While these systems support emotional expression and reflection, they primarily operate within text-based interactions.
They have yet to explore how the emotional content of journal entries might be conveyed through other sensory modalities.
Our study builds on this line of work by introducing a multi-sensory dimension—AI-generated music—that expands journaling’s expressive possibilities and emotional resonance.

\subsection{Music for Mental Health}

Music has long been recognized for its positive effects on emotional states and overall mental well-being. Research has demonstrated that listening to music can yield both physiological and psychological benefits: it may lower heart rate and blood pressure \cite{fallon2020stress, de2022music, burrai2016randomized}, reduce anxiety in medical contexts \cite{brunges2003music, chan2009investigating}, and support emotional regulation by helping individuals manage mood and affective responses \cite{cook2019music, akin2011internet, saarikallio2007role, juslin2008emotional}. 
 
Beyond its physiological effects, music has also been found to serve as a reflective tool for emotional self-regulation. A large body of studies indicates that individuals often use music not only to mirror but also to shift their emotional states \cite{saarikallio2007role, papinczak_young_2015, thoma_emotion_2012}. Specifically, Cook and colleagues found that listeners actively select music to guide themselves toward a desired emotional state or mindset, rather than passively matching their current mood \cite{cook2019music}. 

Building on this perspective, researchers have sought to identify how musical features evoke specific emotional responses.
For example, Meyers \cite{meyers_mood-based_2007}, building on Russell’s circumplex model of emotion \cite{russell_circumplex_1980}, systematically mapped emotions—such as pleasure, arousal, or depression—to musical characteristics like tempo, rhythm, harmony, and volume (See Table \ref{tab:mood-music}).

\begin{table*}[!ht]
  \centering
  \renewcommand{\arraystretch}{1.3}
  \caption{Meyers' mapping of musical features to Russell's circumplex model of emotion. Mode, Harmony, Tempo, Rhythm, and Loudness reflect musical characteristics associated with each mood state.}
  \Description{A table mapping mood states to corresponding musical features based on Meyers’ interpretation of Russell's circumplex model of emotion. Each row represents a mood such as pleasure, excitement, or depression, and lists associated characteristics including musical mode, harmony, tempo, rhythm, and loudness. For example, excitement is associated with major mode, simple harmony, fast tempo, irregular rhythm, and high loudness.}
  \label{tab:mood-music}
  \begin{tabular}{p{2.5cm} p{2.5cm} p{2.5cm} p{2.5cm} p{2.5cm} p{2.5cm}}
    \toprule
    \textbf{Mood} & \textbf{Mode} & \textbf{Harmony} & \textbf{Tempo} & \textbf{Rhythm} & \textbf{Loudness} \\
    \midrule
    Pleasure     & Major   & Simple  & Slow      & Irregular & Medium   \\
    Excitement   & Major   & Simple  & Fast      & Irregular & High     \\
    Arousal      & Neutral & Complex & Very Fast & Regular   & Very High\\
    Distress     & Neutral & Complex & Very Fast & Regular   & High     \\
    Displeasure  & Minor   & Complex & Slow      & Irregular & Medium   \\
    Depression   & Minor   & Complex & Slow      & Irregular & Low      \\
    Sleepiness   & Minor   & Simple  & Very Slow & Irregular & Very Low \\
    Relaxation   & Major   & Simple  & Very Slow & Irregular & Low      \\
    \bottomrule
  \end{tabular}
\end{table*}

In parallel, a growing body of work has explored how music influences writing practices. Studies have shown that music can enhance writing performance and influence the emotional tone of written content, particularly in journaling contexts \cite{ervin2002effect}. For example, Donohoe \cite{donohoe_effect_1999} found that listening to music while writing significantly increased productivity, as measured by word counts. Similarly, Hu \cite{hu_effects_2021} showed that the emotions evoked by music not only influenced how much participants wrote but also shaped the emotional tone of their written content. 

While existing studies have shown that music can enhance writing and emotional expression, they typically rely on curated tracks and focus on outcomes such as productivity or emotional tone, rather than the reflective experience itself.
In particular, little is known about how personalized music, generated in direct response to one’s written emotional expression, might foster a more resonant and meaningful form of reflection.
Recent advances in GenAI models now enable the creation of emotionally aligned, personalized music compositions \cite{hou2022ai}, opening new possibilities for responsive and adaptive journaling experiences.
Our study builds on this potential, exploring how AI-generated music can expand journaling into a multi-sensory, emotionally resonant reflective practice.

\subsection{Potential of AI-Generated Music for Self-reflection}

AI-generated music has a long-standing research history at the intersection of computer science and music studies \cite{roads1985research, ames_cybernetic}.
Early approaches were predominantly rule-based, grounded in composition theory and predefined instructions for machines to follow \cite{hiller1979experimental}.
The introduction of artificial neural networks in the 1980s paved the way for deep learning, with Recurrent Neural Networks (RNNs) gaining attention for their ability to model temporal musical patterns \cite{schmidhuber1992learning}.
More recently, transformer-based models \cite{vaswani_attention} have significantly improved the quality of AI-generated music, enabling the generation of longer, more coherent, and customizable musical sequences \cite{shaw-etal-2018-self}.
These advances have broadened access to music creation, allowing non-experts to produce music aligned with their preferences, contexts, and emotional intentions \cite{dash2024ai}.

Grounded in these advancements, the HCI community has begun exploring AI-generated music as a tool for emotional support, reflection, and self-regulation.
Notably, recent work has integrated generative music into therapeutic contexts, leveraging AI’s adaptability to meet emotional or psychological needs.
For example, Hou \cite{hou2022ai} proposed a theoretical framework that bridges music therapy and AI, using GANs and LSTM models to generate music tailored to user needs and emotional goals.
Similarly, recent work explored co-designing therapist-informed interventions that incorporate generative music, not as technical add-ons but as practically deployable, therapist-centered systems \cite{baglione2021understanding, sun2024understanding}.


In parallel, researchers have worked to expand the infrastructure for emotion-aware music generation by developing datasets that annotate AI-generated music with the prompts and emotions they aim to convey \cite{civit2024sunocaps, doh_lp-musiccaps_2023}. These efforts enable more nuanced emotional alignment and support emerging applications in adaptive music generation.

Overall, this line of research points to the potential of AI-generated music to dynamically respond to users’ emotional states and lived experiences. While personalization has traditionally required manual curation of existing tracks, generative models now make it possible to produce emotionally relevant music on demand \cite{lee2024implementation}. Building on this potential, our study explores how journal-based personalized music generation can deepen emotional engagement and support everyday mental health by integrating AI-driven musical experiences into self-care practices.

\section{Formative Study}
We conducted a formative study with two primary goals: (1) to understand how and why people engage in journaling practices, and (2) to identify design requirements for the system translating journal entries into emotionally resonant music through GenAI. To address these goals, we examined participants’ journaling habits, emotional motivations, and strategies for self-reflection. In parallel, we developed and tested preliminary pipelines that converted written reflections into structured song descriptions using different prompting techniques (See Figure \ref{fig:prompt_pipeline}).

Our formative study involved 15 regular journal writers. Participants were first invited to share recent journal entries and complete a short survey about their journaling habits and emotional goals. They were then presented with multiple AI-generated music pieces based on their entries, each created using a different prompting strategy. For each piece, they rated how well it reflected the emotional tone of their writing and provided qualitative feedback on its resonance and relevance. Further, participants shared their experiences with the system, expectations for emotionally attuned music, and perspectives on how such a tool might enhance journaling.

By analyzing survey responses, music evaluations, and open-ended reflections, we identified key insights that informed both the refinement of our prompt-generation techniques and the design requirements for journaling systems that integrate GenAI music to support self-reflection.

\subsection{Prompts Design}
\begin{figure}
\includegraphics[width=\columnwidth]{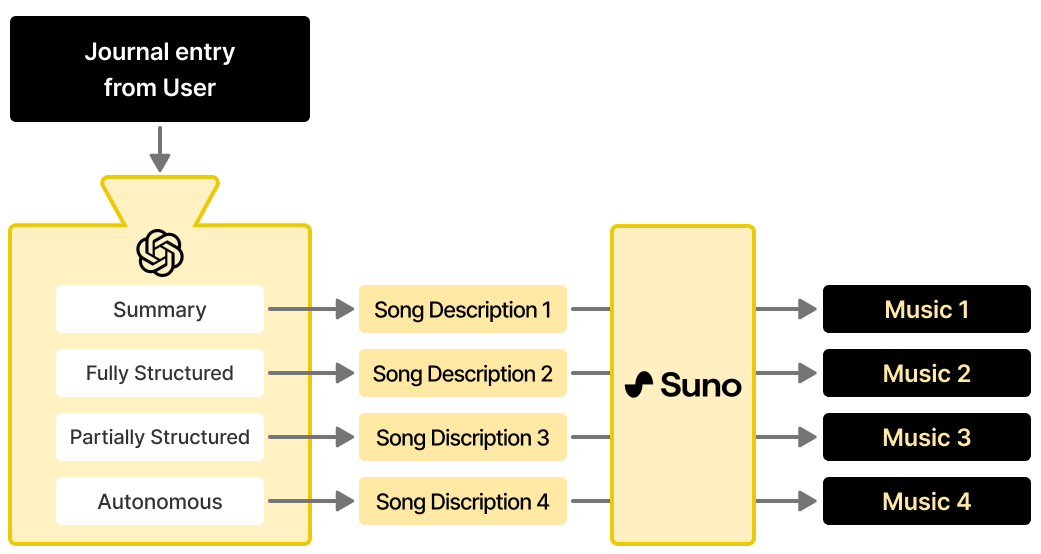}
  \caption{Two-stage pipeline of creating journal-based AI-generated music. GPT-4o analyzes journal entries to create song descriptions using four prompt structures: 1) Summary, 2) Fully Structured, 3) Partially Structured, and 4) Autonomous. Suno converts these descriptions into musical compositions.}
  \Description{The diagram shows a two-stage process for generating music from user journal entries. The first stage involves GPT-4o, which processes the journal entry and generates four types of output. Each of these outputs is transformed into a corresponding song description. The second stage uses Suno, an AI music generation model, to create four unique musical compositions based on these descriptions. Arrows connect the steps in sequential order, visually representing the flow from user input to the final musical output.}
  \label{fig:prompt_pipeline}
\end{figure}

To generate music that well reflects journal entries, we developed a two-stage pipeline utilizing artificial intelligence models (See Figure \ref{fig:prompt_pipeline}). First, we employed GPT-4o \cite{openai_hello_2024}, an LLM that has demonstrated superior performance on the Massive Multitask Language Understanding (MMLU) benchmark \cite{hendrycks_measuring_2021}, to analyze journal entries and generate appropriate song descriptions. These descriptions were then used as input for Suno \cite{suno}, an AI music generation model capable of quickly creating complete musical pieces in various genres based on short written prompts without requiring additional mixing and mastering \cite{dukut_use_2024}.

\begin{figure*}
  \centering
  \includegraphics[width=0.86\linewidth]{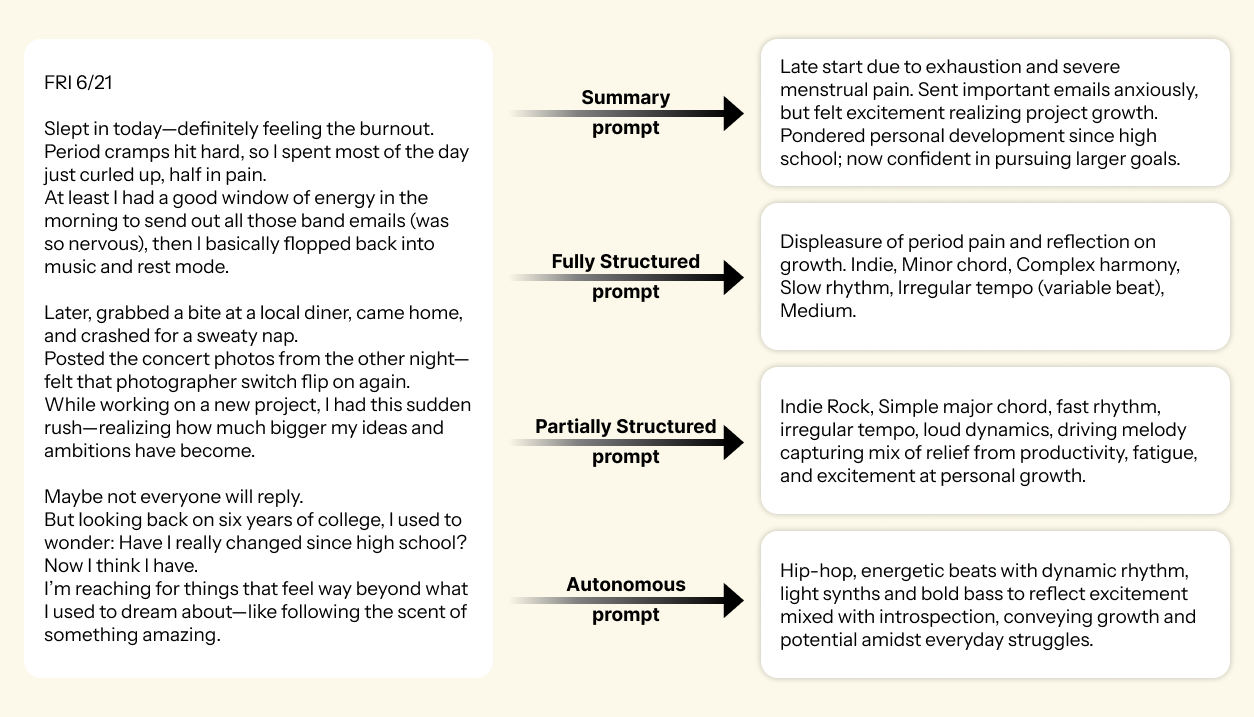}
  \caption{Comparison of song descriptions generated by four different prompts (right) applied to the same journal entry written by participant F5 (left). The outputs show how each prompt format guided the generation of each music description.}
  \Description{A figure showing a journal entry written by participant F5 on the left, and four song descriptions on the right, each generated using a different prompt format: Summary, Fully Structured, Partially Structured, and Autonomous. The descriptions vary in structure and detail, reflecting the influence of each prompt type on the output.}
  \label{fig:promptSample}
\end{figure*}

The development of our prompt system for GPT-4o was fundamentally based on Meyers' framework \cite{meyers_mood-based_2007}, which identified specific musical elements that effectively express different emotional states based on their valence and arousal levels in Russell's circumplex model of emotion \cite{russell_circumplex_1980} (See Table \ref{tab:mood-music}). Building upon these established emotion-music relationships, we designed four distinct prompts: \textbf{Summary}, \textbf{Fully Structured}, \textbf{Partially Structured}, and \textbf{Autonomous}. Each prompt reflects a different balance of musical-emotional mappings, incorporation of journal context and narrative elements, and degree of LLM autonomy (See Table \ref{tab:prompt-structures} and sample outputs in Figure \ref{fig:promptSample}).

\begin{itemize}
    \item \textbf{Summary prompt} preserved the journal’s narrative by condensing it into a short sentence (under 200 characters, as required by Suno AI at the time of the study). It began with a music genre and highlights the central emotion and event. No emotion–music rules were applied, but the LLM operated with moderate autonomy to distill the entry into a concise, emotionally expressive format.  
    \item \textbf{Fully Structured prompt} required the LLM to first extract the dominant emotion from the journal entry and place it at the beginning of the song description. It then applied predefined mappings between that emotion and five musical elements—chord, harmony, rhythm, tempo, and loudness—based on Meyers’ framework \cite{meyers_mood-based_2007}.     
    \item \textbf{Partially Structured prompt} aimed to balance emotional mapping rules with narrative expression. It drew on the same emotion–music framework but allowed more flexible phrasing and partial incorporation of the journal’s tone and content. 
    \item \textbf{Autonomous prompt} gave the LLM high autonomy, requiring only the inclusion of a genre. This allowed the LLM to freely interpret the emotional tone and narrative context of the journal entry, with no predefined rules or mappings applied. 
\end{itemize}


Through this pipeline, journal entries were first processed by GPT-4o to generate song descriptions, which were then transformed into musical pieces using Suno. We used these AI-generated compositions in our formative study to explore how users perceived and responded to music derived from their own journal entries.

\begin{table*}[!h]
  \setlength{\tabcolsep}{3pt}
  \renewcommand{\arraystretch}{1.1}
  \centering
  \caption{Comparison of prompt structures for journal-based song description generation. The (Prompt) column denotes four different prompt types. The (Emotion-Music Rules) column indicates the degree to which Meyer's mapping of musical features to emotion was enforced in the prompt. The (Journal Narrative) column denotes how much of the original journal entry's narrative was incorporated. The (AI Autonomy) reflected the level of creative freedom given to the LLM.}
  \Description{A table comparing four types of prompts used for generating music descriptions from journal entries: Summary, Fully Structured, Partially Structured, and Autonomous. Each prompt type is evaluated across three criteria—Emotion-Music Rules, Journal Context, Narrative, and AI Autonomy.}
  \label{tab:prompt-structures}
  \newlength{\remainingwidth}
  \newlength{\mycolwidth}
  \setlength{\remainingwidth}{\dimexpr\textwidth - 4cm - 8\tabcolsep\relax}
  \setlength{\mycolwidth}{\dimexpr\remainingwidth / 4\relax}
  \begin{tabular}{
      p{4cm}
      *{4}{>{\centering\arraybackslash}p{\mycolwidth}}
    }
    \toprule
      & \textbf{Summary}
      & \makecell{\textbf{Fully Structured}}
      & \makecell{\textbf{Partially Structured}}
      & \makecell{\textbf{Autonomous}} \\
    \midrule
    \makecell[tl]{\textbf{Emotion‑Music Rules}}
      & None
      & \makecell{Strictly enforced\\predefined rules}
      & \makecell{Balance of rules\\and flexibility}
      & None \\
    \addlinespace
    \makecell[tl]{\textbf{Journal Narrative}}
      & Fully preserved
      & Minimum
      & Partially incorporated
      & Fully incorporated \\
    \addlinespace
    \makecell[tl]{\textbf{AI Autonomy}}
      & Middle
      & Low
      & Middle
      & High \\
    \bottomrule
  \end{tabular}
\end{table*}

\subsection{Formative Study Participants}
A total of 15 participants (3 male, 12 female; \textit{Mean} = 24 yrs, \textit{min} = 20 yrs, \textit{max} = 26 yrs) were recruited for the formative study through postings on university online communities. All participants were native Korean speakers who reported maintaining regular journaling practices and writing diaries at least twice a week during the month prior to the study. Participant details are provided in Appendix~\ref{appendix:formativeparticipant}. The research procedures were approved by the Institutional Review Board (IRB) of Seoul National University, where the study was hosted. The study goals and procedures were thoroughly explained to each participant, and written informed consent was obtained. Participants received 15,000 KRW (approx. 10 USD) for their participation. 

\subsection{Procedure}

\subsubsection{Pre-study Survey}
Prior to the formative study, we conducted an online survey to collect demographic information and details about participants’ journaling habits. Participants were then asked to submit two journal entries that reflected contrasting emotional valence: one expressing pleasant emotions and another expressing unpleasant emotions.


\subsubsection{Formative user study session}
We conducted 60-minute user study sessions either in person or remotely via Zoom. Prior to each session, we obtained participants’ consent for audio recording and for the use of their submitted journal entries for research purposes. 
One researcher led the interview while another took notes.

Each session began with open-ended questions about participants’ journaling practices and their use of music in daily life. We asked about their motivations for journaling and whether they used music as a tool for emotional regulation.

In the second phase, participants were presented with AI-generated music pieces created from their submitted journal entries. 
For each entry, we generated four versions of music using different prompting approaches (see Table \ref{tab:prompt-structures}). All pieces were instrumental to minimize confounding effects from lyrics and ensure that participants focused on the musical elements themselves.

After listening to each piece, participants rated its emotional and narrative reflectiveness on a 7-point Likert scale (1 = “not reflective at all,” 7 = “highly reflective”). They were then asked to explain the reasoning behind their ratings and to reflect on how well the music captured the tone or meaning of their writing.

Finally, participants shared their experiences with the AI-generated music derived from their journal entries, their expectations for such music pieces, and their perspectives on how such tools could enhance their journaling experience.

\subsection{Analysis}
Our analysis combined quantitative and qualitative methods to offer a comprehensive understanding of participants’ experiences. For the quantitative analysis, we used descriptive statistics to summarize participants’ survey responses. To assess differences in reflectiveness across the four prompting approaches, we first tested for normality using the Shapiro-Wilk test, which indicated significant non-normality. Based on this result, we employed the Mann-Whitney U test to compare reflectiveness ratings between prompting conditions.

For the qualitative analysis, we conducted thematic analysis of the interview transcripts \cite{saldana_coding_2024}. Each researcher first conducted open coding independently, then collaboratively reviewed and refined the codes through iterative discussions. This process resulted in a set of overarching themes that captured key patterns across participant responses and informed emerging design requirements.

\subsection{Findings}
In this section, we outline the main findings of the formative study in five parts. Each finding is supported by participant interviews and survey data, with participants referred to by their assigned identification numbers (e.g., F1, F2).

\subsubsection*{\textbf{Finding \#1: Participants projected journaling’s emotional and archival value onto AI-generated music}}
Participants described two primary benefits of journaling: emotional regulation and record-keeping. They stated that journaling helped them manage emotions and reduce stress through the act of articulating and examining their experiences. At the same time, many valued their journals as personal archives—tools for capturing memories and supporting reflection over time. While some participants emphasized either emotional processing or record-keeping, most actively engaged in both. Importantly, many expected these benefits to extend to journal-based AI-generated music. F5 noted, ``\textit{I think journal-based AI-generated music would help maintain pleasant emotions in an energetic state while helping to resolve unpleasant emotions, just like journaling does.}''


\subsubsection*{\textbf{Finding \#2: Participants valued emotional and narrative alignment in journal-based AI-generated music}}

Participants consistently found journal-based AI-generated music more satisfying when it effectively captured both the emotional tone and narrative content of their writing. F4 noted that she ``\textit{valued music that closely reflected the journal’s emotional content more than music that simply matched their personal preferences}.'' 

When evaluating this reflectiveness with the journals and generated music pieces, participants attended to musical elements such as instrumentation, tempo, melody, and transitions. They assessed how well these features matched the emotions and events described in their journals. For instance, F10 noted, ``\textit{The emotions of that day matched better with music containing synthesizers rather than acoustic instruments.}'' This alignment between music and lived experience helped participants revisit and reprocess their emotional states. F15 noted, ``\textit{Journal-based AI-generated music could serve as an opportunity to look back and reflect on the emotions felt at the time of writing the journal.}''

\begin{figure}[t]
  \centering
  \includegraphics[width=\columnwidth]{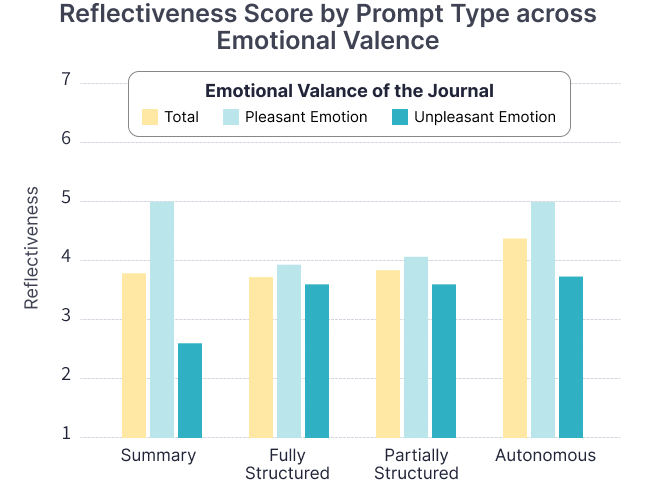}
  \caption{Mean reflectiveness scores (1–7 Likert) for music generated from journals written with four prompt types. }
  \Description{Grouped bar chart displaying reflection scores grouped by prompt type—Summary, Fully Structured, Partially Structured, and Autonomous. The x-axis lists four prompt types—Summary, Fully Structured, Partially Structured, and Autonomous—while the y-axis runs from 1 (low reflectiveness) to 7 (high reflectiveness). For each prompt type three adjacent bars appear: the first bar representing the overall mean reflection score, the second bar for entries expressing pleasant emotion, and the third bar for entries expressing unpleasant emotion. In every prompt group the pleasant entries scores are higher than the unpleasant entries. This consistent pattern shows that music generated from unpleasant-emotion journals was perceived as less reflective—regardless of prompt structure—than music generated from pleasant-emotion journals.}
  \label{fig:boxplot_formative}
\end{figure}

Participants’ evaluations of musical reflectiveness were also influenced by the emotional valence of their journal entries. A Mann-Whitney U test revealed that the reflectiveness scores for music based on entries expressing unpleasant emotions (\textit{Mean} = 3.63, \textit{SD} = 2.03) were significantly lower than those based on entries expressing pleasant emotions (\textit{Mean} = 4.59, \textit{SD} = 1.72; W = 1825, \textit{p}<.01) (see Figure \ref{fig:boxplot_formative}). Participants explained that negative emotions were more difficult to translate into music and required greater nuance to feel accurate:  ``\textit{In the case of unpleasant emotions, the music needs to match even the subtle details for it to feel like it accurately reflects those emotions}'' (F4).

\subsubsection*{\textbf{Finding \#3: Participants wanted music that could support their emotional regulation goals}}
Although participants liked music that reflected their journal content (\textit{Finding \#2}), their preferences were also shaped by how they wanted to regulate their emotions. Most aimed to move toward a pleasant, low-arousal state, and their musical preferences shifted depending on whether they were processing positive or negative emotions. Participants commonly preferred music that helped soothe or shift away from unpleasant emotions, while maintaining or amplifying pleasant ones. Rather than consistently preferring music that matched the emotional tone of their journal entries, they often chose music that aligned with their regulation goals. For example, F4 stated, ``\textit{When I want to moderate my emotions, I'd like to listen to music with opposite emotions, and when I want to maintain them, I'd like to play music with a similar mood.}''

\subsubsection*{\textbf{Finding \#4: Participants rated music generated by the autonomous prompt as the most reflective}}
The statistical analysis of the reflectiveness scores showed that participants rated music generated from the autonomous prompt highest in reflectiveness (\textit{Mean} = 4.37, SD = 1.77), and music from the fully structured prompt lowest (\textit{Mean} = 3.83, SD = 1.90) (see Figure \ref{fig:boxplot_formative}). This indicated that granting the AI greater creative autonomy, rather than relying on strictly predefined emotion–music mappings, may result in outputs that participants found more personally resonant.



\subsubsection*{\textbf{Finding \#5: Participants exhibited diverse preferences for control and customization in journal-based AI music generation}}
Participants expressed varied preferences for how much control they wanted over the AI music generation process. Some of them wanted to adjust specific musical elements to match their preferences: ``\textit{I want to incorporate my preferred genre and tempo into the music}'' (F10). Others sought modifications to enhance emotional expression and support their regulation strategies: ``\textit{I want to modify the music to better express my feelings}'' (F3). Some participants requested more control over the generation process itself: ``\textit{I would like to generate multiple pieces and choose the one I like best}'' (F12). Several participants also wished for lyrics to deepen emotional resonance: ``\textit{I think having lyrics would help me focus better on my emotions}'' (F15).


However, participants differed in how much direct control they desired. While F2 wanted more user intervention (``\textit{I wish there were more opportunities for active user intervention}''), F8 preferred a more automated experience (``\textit{I would like the AI to generate music automatically without any additional modifications}''). Notably, no participants expressed interest in writing their own song descriptions from scratch, indicating a desire for lightweight interaction without requiring extensive creative effort.


\subsection{Design Requirements} \label{subsection:design-requirements}
Our findings revealed that participants' existing journaling practices strongly influenced their expectations for journal-based AI-generated music. Participants sought music that actively supported their emotional regulation goals and judged its effectiveness based on how well it reflected the emotional and narrative content of their journals. Among different prompting approaches, autonomous prompts were rated most highly, suggesting that creative flexibility improved emotional resonance. Participants also expressed varied preferences for user control over music generation, though notably, none wanted to write song descriptions from scratch.

Drawing from these insights, we propose the following design requirements for systems that integrate AI-generated music into journaling practices:



\subsubsection*{\textbf{Requirement \#1: Preserve and amplify the emotional and archival functions of journaling} (Finding \#1)}
The system should support the two core functions that users associate with journaling: emotional processing and personal archiving. These existing practices shaped how participants approached AI-generated music, with many expecting it to offer similar emotional and reflective benefits. To meet these expectations, the system should help users both revisit past experiences through music and engage with their emotions in meaningful ways. Importantly, it should flexibly accommodate different journaling styles, whether users focus more on emotional expression, memory preservation, or both.


\subsubsection*{\textbf{Requirement \#2: Generate music that reflects journal content using autonomous prompts} (Findings \#2, \#4)}
The system should prioritize generating music that captures both the emotional tone and narrative context of journal entries. Participants consistently evaluated AI-generated music based on how well it reflected the content of their writing. Among the different prompting strategies, autonomous prompts produced the most reflective outputs. By giving the LLM more interpretive freedom, this approach enabled nuanced translations of journal entries into music while preserving the emotional integrity of the original journal entries.



\subsubsection*{\textbf{Requirement \#3: Support users' emotional regulation goals} (Finding \#3, \#5)}
The system could facilitate users' preferred emotional regulation strategies, rather than simply mirroring the emotions expressed in their journal entries. This adaptability will be important as participants exhibited diverse preferences in emotional regulation: some aimed to transform unpleasant emotions into more positive states, while others sought to maintain or gently moderate pleasant emotions. To accommodate these needs, the system should allow users to specify their desired emotion regulation goals and generate music that aligns with their intention.

\subsubsection*{\textbf{Requirement \#4: Enable flexible user control over music generation} (Finding \#5)}
The system should support flexible user control throughout the music generation process, addressing participants’ varied preferences for involvement. This can be achieved across two key phases:

\paragraph{Pre-generation:} The system should allow users to specify parameters such as their desired emotional regulation strategy, preferred genre, or inclusion of lyrics. Supporting these preferences ensures that users can shape the emotional and stylistic direction of the music before it is generated.

\paragraph{Post-generation:} The system should provide multiple variations of generated music and allow users to select the version that best fits their intent. It should also offer lightweight editing tools for adjusting musical elements such as tempo, instrumentation, or mood. These options can help users fine-tune the output to better reflect their journal content and emotion regulation goals.



\subsubsection*{\textbf{Requirement \#5: Maintain AI autonomy in music generation} (Findings \#4, \#5)}
The system should leverage AI autonomy in the music generation process while accommodating varying user preferences for control. Our findings showed that music generated through autonomous prompts, where the AI had greater freedom to interpret journal entries, received higher ratings across several criteria. While participants expressed varying desires for control, none preferred to manually direct the full generation process. Maintaining AI autonomy in interpreting journal content and making musical decisions ensures expressive, reflective outputs, while allowing optional user input supports personalization without increasing the burden. This balance can help sustain the creative advantages of GenAI while respecting users' desired levels of control over the generation process.


\section{NoRe: Note \& Rest}
\begin{figure*}
  \includegraphics[width=0.65\textwidth]{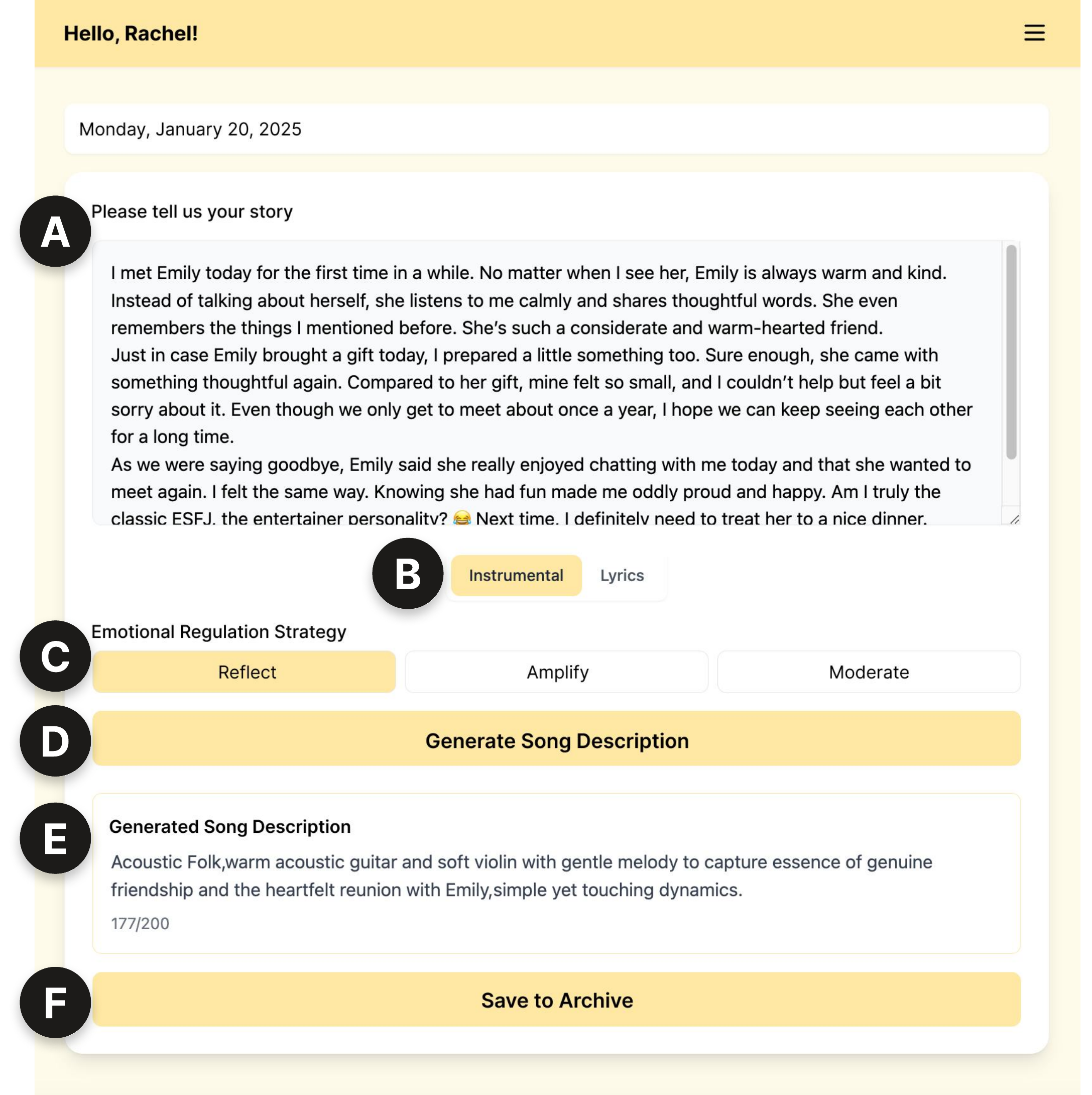}
  \caption{NoRe interface. (A) The main text input area for journal writing. (B) Option toggle for lyrics inclusion. (C) Emotional regulation strategy options (Reflect, Amplify, Moderate). (D) Song description generation button. (E) Generated song description. (F) Save button.}
  \label{fig:system}
  \Description{The figure shows the NoRe web interface for journal-based music generation. At the top is a text editor area displaying a personal journal entry about meeting a friend named Emily. Below the text editor are several control elements: a toggle switch between Instrumental and Lyrics options, three emotional regulation strategy buttons (Reflect, Amplify, Moderate), and a button to generate song descriptions. The bottom section shows a generated song description suggesting acoustic folk music with warm guitar and soft violin to match the journal's emotional tone, followed by a Save to Archive button. The interface uses a warm, cream-colored theme with clear visual hierarchy and intuitive controls.
 }
\end{figure*}

\begin{figure*}[!t]
  \includegraphics[width=0.75\textwidth]{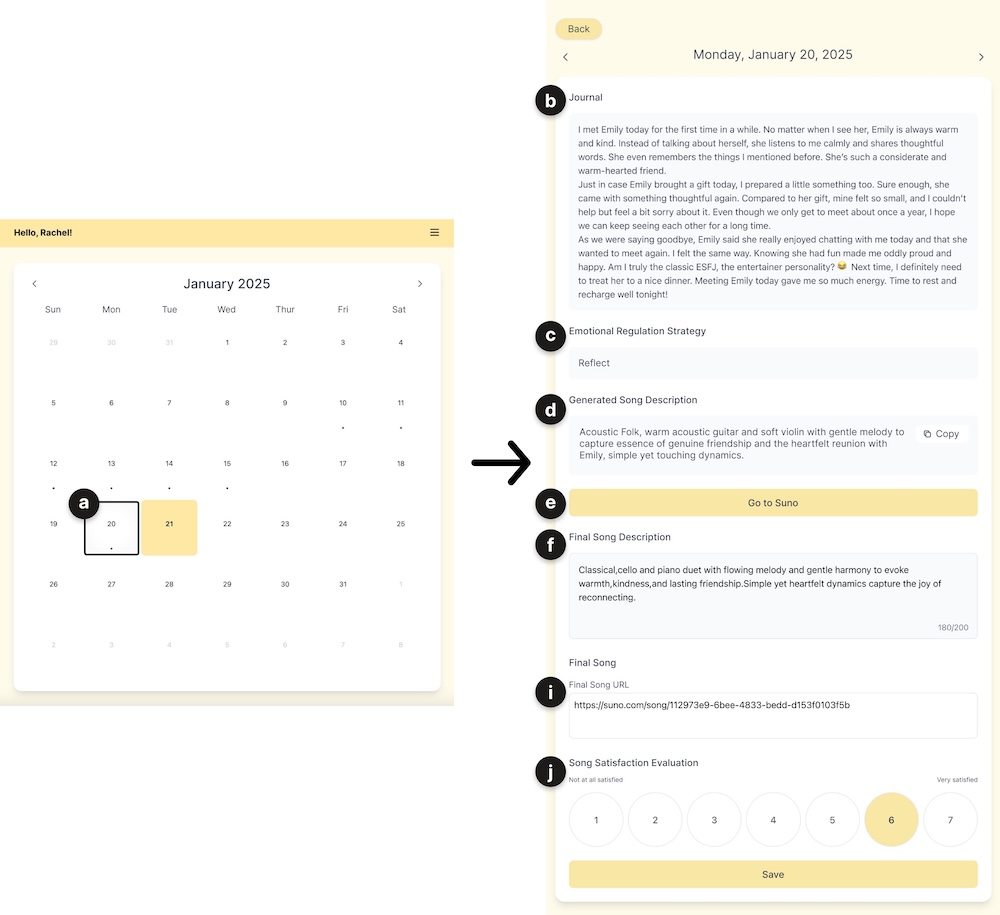}
  \caption{NoRe archiving interface. (a) Calendar interface showing archived dates with dots. Clicking on it navigates to the archiving page (b) the journal entry view. (c) Selected emotional regulation strategy. (d) Initial song description (generated). (e) Button to Suno. (f) Final song description after iterations. (i) Generated song URL section for archival. (j) Song satisfaction evaluation using a 7-point Likert scale.}
  \label{fig:archive}
  \Description{The figure shows NoRe's archive interface split into two main sections. The left side displays a calendar view for January 2025 with dots indicating dates where entries exist. The right side shows the detailed view of a selected entry, including the original journal text, emotional regulation strategy selected (Reflect), initial and final song descriptions suggesting acoustic folk and classical arrangements, a link to the generated music, and a 7-point satisfaction rating scale. The interface uses a clean, organized layout with clear navigation between calendar and entry views.}
\end{figure*}

Informed by the findings and design requirements derived from the formative study, we designed and developed \textbf{NoRe}, an abbreviation of Note \& Rest. NoRe is a web-based application that allows users to journal and generate AI-generated music based on their journal entries. In addition, NoRe enables users to archive their journal entries and AI-generated music and revisit them. The interface can be found in Figure \ref{fig:system} and \ref{fig:archive}.

\subsection{Features}
NoRe's key features were designed to align with the design requirements outlined in Section \ref{subsection:design-requirements}. First, NoRe adopts a familiar text editor interface for journal entry creation to preserve and enhance users' established journaling practices \textbf{(Design Requirement \#1)}. 
In addition, NoRe supports comprehensive archiving of both journal entries and their corresponding musical outputs, reinforcing the practice of revisiting and reflecting on past experiences \textbf{(Design Requirement \#1)}.


In addition, NoRe includes a feature that allows users to align the generated music with their emotional regulation goals \textbf{(Design Requirement \#2)}. Users can specify how they want the music to respond to the emotions expressed in their journal entry by selecting one of three options: \textbf{maintain}, \textbf{amplify}, or \textbf{moderate}.

Further, NoRe provides pre-generation customization through a set of user-adjustable options \textbf{(Design Requirement \#3)}. Users can specify whether to include lyrics in the generated music. As mentioned above, they can also choose an emotional regulation strategy (maintain, amplify, or moderate). Based on these selections, the system automatically generates an initial song description, which serves as input to the AI music generation model \textbf{(Design Requirement \#5)}. However, users can modify this description to better reflect their intent and musical preferences.

Finally, to support user agency in the post-generation phase \textbf{(Design Requirement \#4)}, NoRe produces two distinct music versions for each journal entry, allowing users to select the composition they prefer to archive alongside their journal entry.

\subsection{Implementation}
NoRe was developed as a web application. The front-end was developed using Next.js \cite{vercel_nextjs_2025}. For back-end development, we utilized MongoDB \cite{mongodb_2024} to store and manage user data, including journal entries, emotional regulation strategies, song descriptions, generated music, and corresponding user ratings. Secure user authentication was also implemented using JSON Web Tokens (JWT) \cite{auth0_jwt_2025}, enabling encrypted login functionality. The application was deployed via Vercel \cite{vercel_vercel_2025}.

Song descriptions were generated using OpenAI’s GPT-4o \cite{openai_hello_2024}, the state-of-the-art model available at the time of the study, based on journal content and users’ pre-generation preferences. Since Suno \cite{suno}, the music generation model, did not provide an API, participants were required to access the Suno platform separately to generate music using these song descriptions. To support this workflow, each participant was provided with a dedicated Suno account.


\subsection{Example User Scenario}
To illustrate how users interact with NoRe, we present an example user scenario: Upon logging into NoRe, users are directed to the main screen, where a journal editor is immediately available for composing a new entry. After completing their writing, users scroll to the ``Music Generation Settings'' panel at the bottom of the screen.

Here, users select one of three emotional regulation strategies (maintain, amplify, or moderate) and indicate whether they prefer instrumental music or lyrics using a toggle button. Once preferences are set, clicking the ``Generate Song Description'' button prompts NoRe to create a song description based on the journal content and selected settings.

The generated song description appears in the interface alongside a copy button. Users can revise the description if desired, then copy it and follow the provided link to the Suno website. There, they configure the instrumental setting to match their earlier selection, paste the prompt, and initiate music generation.

Suno produces two musical pieces by default. Users listen to both and may optionally revise the prompt and regenerate the music. Once they select their preferred track, they copy the final song description and the track’s link, return to NoRe, and paste them into the corresponding fields. They then rate the chosen music on a 7-point Likert scale and click ``Save'' to archive the journal entry, selected music, and rating.

To access previous entries and music, users navigate to the archive page by clicking the ``Archive'' button in the top-right navigation bar. The archive page displays a calendar view, with dates containing journal entries and music marked by dots. Clicking a marked date allows users to revisit the journal content and re-experience the associated music.

\section{In-the-Wild Evaluation of NoRe}
To explore the potential of augmenting journaling experiences with AI-generated music, we conducted a 7-day in-the-wild evaluation study of NoRe with 15 regular journal writers. Through analysis of both qualitative and quantitative data, including system usage logs, we gained insights into how participants perceived and engaged with journal-based AI-generated music, and its implications for mental well-being and self-reflection.

\subsection{Procedure}
The in-the-wild study consisted of three phases: 1) Pre-deployment, 2) Deployment, and 3) Post-deployment.

\subsubsection{Pre-deployment Phase (30 min)}
We first invited each participant to a remote onboarding session via Zoom. During each session, one researcher led the conversation while the other researcher took detailed notes.
We introduced the study goals and procedures, explaining how participants would engage with journaling and music listening. Subsequently, we gave a live demonstration of how to use NoRe. To ensure participants could reference the system's features at any time, we provided participants with a comprehensive how-to document for NoRe. The session took about 30 minutes.

\subsubsection{Deployment Phase (7 days)} \label{evaluation questions}
Participants began using NoRe the day after the onboarding session and were instructed to engage with the system for at least three of the seven days. Compensation was not tied to the frequency of use in order to minimize bias.

After each journaling session and interaction with the AI-generated music, participants completed an online survey consisting of 7-point Likert scale questions assessing how well the music reflected their journal content, the emotional benefits of the experience (e.g., reminiscence, emotional regulation, and self-understanding), and the usability of the NoRe system that day. To evaluate these dimensions, we organized eight survey items into three categories: (1) \textbf{journal-to-music reflectiveness}, (2) \textbf{system usability}, and (3) \textbf{emotional benefits}. Below are the items presented to participants:

\paragraph{\textbf{Journal-to-music Reflectiveness}}
\begin{description}
\item[Q1:] ``The journal-based music generated today adequately captures the day's significant events.''
\item[Q2:] ``I think the journal-based music generated today sufficiently represents the day's important emotions.''
\item[Q3:] ``I think the journal-based music generated today adequately reflects my main thoughts from today.''
\item[Q4:] ``I feel the journal-based music generated today genuinely captures my true self.''
\end{description}


\paragraph{\textbf{System Usability}}
\begin{description}
\item[Q5:] ``Today's journaling and music generation process  was smooth''
\end{description}


\paragraph{\textbf{Emotional Benefits}}
\begin{description}
\item[Q6 (Reminiscence):] ``Today's use of the NoRe helped me remember the day's events''
\item[Q7 (Emotional regulation):] ``Today's use of the NoRe aided in regulating my emotions''
\item[Q8 (Self-understanding):] ``While using the NoRe today, I gained a deeper understanding of myself and my emotions''
\end{description}

In addition to the survey items described above, participants were also asked to rate their \textbf{satisfaction} with each piece of generated music directly within the NoRe interface (7-point Likert scale). Throughout the study period, we collected \textbf{system usage logs}, including journal entries, initial and finalized song descriptions, selected emotional regulation strategies, lyric preferences, and song URLs. All data were securely stored in our database.


\subsubsection{Post-deployment Phase (60 min)}
After the 7-day deployment period, we conducted follow-up semi-structured interviews with all participants. The interviews focused on three key topics: (1) overall user experience and usability, (2) perceived effectiveness of journal-based AI-generated music, and (3) future suggestions for journal-based AI music systems. All interviews were conducted remotely via Zoom and lasted approximately 60 minutes each.

\subsection{Participants}
A total of 15 participants (2 male, 13 female; \textit{Mean} = 24.07 yrs, \textit{range} = 19-29 yrs) were recruited for the in-the-wild evaluation study. As in the formative study, all participants were native Korean speakers who reported maintaining regular journaling practices, writing journals at least twice a week during the month prior to the study. While recruiting for the evaluation study, we contacted the participants from the formative study to ask for additional involvement in the research. Overall, six participants from the formative study agreed to participate in the evaluation study. The remaining participants were newly recruited through postings on university-affiliated online communities. Detailed information about participants is in Appendix~\ref{appendix:evaluationparticipant}.

The research procedures were approved by the Institutional Review Board (IRB) of Seoul National University. The study goals and procedures were thoroughly explained to each participant, and written informed consent was obtained. We compensated all participants with 50,000 KRW (approx. 35 USD) for their involvement in the in-the-wild evaluation study. 



\subsection{Analysis}
Our analysis of the in-the-wild evaluation employed both quantitative and qualitative methods. For the quantitative analysis, the unit of analysis was each journal entry–generated music pair, rather than individual participants. We primarily performed descriptive statistical analyses to summarize Likert-scale responses regarding music reflectiveness, emotional benefits, system usability, and satisfaction with each generated music piece.

In addition to descriptive summaries, we conducted comparative analyses for behaviorally distinct conditions (e.g., whether the song description was modified or not). Given the non-normal distribution of the data, we used Mann-Whitney U tests to compare reflectiveness and satisfaction scores across these conditions.

To further explore behavioral patterns, we employed chi-squared tests to examine associations between participants’ journal content and their interaction choices, such as lyric inclusion, emotional regulation strategy selection, and prompt modification. For example, we analyzed how the sentiment of a journal entry (e.g., pleasant vs. unpleasant) was associated with the selected emotion regulation strategy or lyrics preference.


For the qualitative data from the post-deployment interview, we conducted a thematic analysis  \cite{saldana_coding_2024}. Each research team member independently conducted open coding of interview transcripts, followed by collaborative review sessions to cross-examine the codes. Through iterative discussions, we developed a set of themes that captured participants’ experiences, perceptions, and reflections on using journal-based AI-generated music. Representative quotes were selected to illustrate key findings.

\subsection{Ethical Considerations}
Although this study received IRB approval, we recognize the ethical complexities of conducting in-the-wild research involving highly personal content such as journals.  To safeguard participants’ privacy, we took several precautions. During the onboarding session, we clearly explained the purpose of data collection and how the data would be stored and analyzed. Journal data from participants who declined this use were excluded from all analyses. Furthermore, participants retained the right to request deletion of their journal entries at any time.


\section{Results}
\begin{table*}[b]
\caption{Overview of Journal Entry Sentiments and Corresponding Emotion-Regulation Strategies}
\Description{A table summarizing features of 50 journal entries and their use in music generation. It shows the sentiment distribution—neutral, pleasant, unpleasant, and mixed—and the number and percentage of entries using lyrics. It also reports emotion-regulation strategies applied to each entry: amplify, maintain, moderate, and modified. Most entries were pleasant or mixed in sentiment, with 'maintain' as the most common emotion-regulation strategy.}
\label{tab:entryOverview}
\centering
\begin{tabular}{lrrrrrrrrrr}
\toprule
~                  &  & \multicolumn{4}{c}{Sentiment of the entry} &     \multicolumn{3}{c}{Emotion-regulation strategy} &  \\
\cmidrule(lr){3-6} \cmidrule(lr){7-10}
~                  & entry N & neutral & pleasant & unpleasant & mixed  & amplify & maintain & moderate \\
\midrule
Count                &        50 &       9 &       17 &       11 &    13 &     11 &    30 &     9 \\
Ratio               &      100\% &   18\% &    34\% &    22\% & 26\%  & 22\% & 60\% & 18\% \\
\bottomrule
\end{tabular}
\end{table*}

The results of the in-the-wild evaluation are separated into three sections: 1) Overall reflectiveness and satisfaction of AI-generated songs based on journal entries, 2) User Behaviors and Preferences in Music-Augmented Journaling, and 3) Perceived Effectiveness of NoRe. Each finding is supported by participant interviews and quantitative data, with participants referred to by their assigned identification numbers (e.g., P1, P2)

\subsection{Overall Reflectiveness and Satisfaction with AI-generated Songs}
To evaluate how well the music reflected participants' journal entries, we collected responses to four journal-to-music reflectiveness questions (Q1–Q4) after each journaling session. These items were designed to assess different aspects of how participants perceived their written experiences being translated into music. Given the high internal consistency among the items (Cronbach’s $\alpha$ = 0.87), we averaged the scores to calculate a composite reflectiveness score. In addition to reflectiveness, we also analyzed participants’ satisfaction with the generated song they ultimately chose to archive alongside their journal entry.

The results showed moderately high scores across both metrics. Participants reported that the generated music reflected their journal entries reasonably well, with an average reflectiveness score of 5.50 out of 7 (SD = 1.21). They also reported relatively high satisfaction with the music, averaging 5.52 out of 7 (SD = 1.11). These findings suggest that participants often perceived a meaningful alignment between their written reflections and the generated music, and that they found the listening experience both personally engaging and emotionally resonant. Together, these results indicate that NoRe’s two-stage pipeline was generally effective in supporting reflective and satisfying journaling experiences.

In the follow-up interviews, participants consistently noted that accurate reflection of their journal entries strongly influenced their satisfaction with the generated music, reinforcing \textbf{\textit{Finding \#2}} from the formative study. For instance, P12, who rated the journal-to music reflectiveness a 7 out of 7, remarked: ``\textit{The baseline and top notes sounded great, which was fascinating. Today’s memories were so happy that I wanted the music to bring back this day when I listened to it later. So, I chose an upbeat piece that best expressed my journal.}'' This underscores how musical representation of journal content played a central role in shaping participants’ positive evaluations.

Participants also emphasized the importance of alignment between the music and their selected emotion-regulation strategies, reflecting \textbf{\textit{Finding \#3}} from the formative study. P14, who selected the ``maintain'' strategy and gave a satisfaction rating of 7, noted: ``\textit{The music seemed to genuinely capture my emotions. Since it was created based on my feelings, I liked that it didn’t feel like the song was intentionally trying to amplify or suppress my emotions.}'' Similarly, P4, who chose the ``moderate'' strategy for a negatively toned journal entry, said: ``\textit{The slightly more vibrant feel of the music seemed to refresh my mood, which I liked},'' rating it highly satisfactory. 

Together, these findings underscore the importance of aligning musical output with users’ written reflections and emotional goals. Participants’ responses suggest that NoRe’s design was largely effective in supporting these needs.

\subsection{User Behaviors and Preferences in Music-Augmented Journaling}

Our study collected a total of 50 journal entries from 15 participants over a 7-day period. On average, participants wrote 3.33 journal entries over the 7-day study period (SD = 0.79), with the most active participant submitting six entries.

Through follow-up interviews, we asked participants to indicate the major emotional content of each entry (see Table \ref{tab:entryOverview}). Of these, 18\% (9 entries) did not express significant emotional content, while the remaining 82\% (41 entries) did. Among the entries with expressed emotions, 41.5\% (17 entries) conveyed primarily pleasant sentiments (such as \textit{``happiness''} and \textit{``fun''}), 26.8\% (11 entries) expressed unpleasant sentiments (such as \textit{``worried''} and \textit{``sad''}) and 31.7\% (13 entries) demonstrated mixed emotions of both positive and negative sentiments.

Based on system logs from the music generation process, 40\% of journal entries (20 out of 50) were paired with music that included lyrics in the final versions participants chose to archive. Emotion-regulation strategy selections showed a strong preference for maintaining one’s emotional state, accounting for 60\% of archived entries (30 out of 50). 

In addition, in 30\% of entries (15 out of 50), participants modified the initial song prompt before selecting the final version. In total, 46.7\% of participants (7 out of 15) modified at least one prompt during the study period, indicating active engagement with the music generation process.

Overall, participants reported a positive user experience with NoRe. The usability item (Q8), assessed after each use, yielded an average score of 6.32 out of 7 (SD = 0.93), suggesting that participants consistently found the system smooth and easy to use. In follow-up interviews, most participants noted that journaling with NoRe felt similar to their usual journaling routines. This suggests that, for many users, the integration of music generation may not have disrupted existing habits or imposed noticeable additional cognitive effort, particularly among those already comfortable with reflective writing practices.

In the following section, we report findings on how participants engaged with \textbf{NoRe’s key customization features}, including \textbf{song description modification}, \textbf{emotion-regulation strategy selection}, and \textbf{lyrics inclusion}.

\subsubsection{Song Description Modifications}

Both the music’s reflectiveness and satisfaction scores were significantly associated with whether participants modified the song description before finalizing the generated piece. To examine this effect, we conducted Mann-Whitney U tests comparing scores between tweaked and non-tweaked song generations (see Table~\ref{tab:tweak_score}). 

The results revealed significant differences in both reflectiveness (Mann-Whitney \textit{U} = 157, \textit{p} = .025) and satisfaction (\textit{U} = 143, \textit{p} = .008) scores, with higher median scores observed for tweaked prompts. These findings suggest that participants who adjusted the song description were more likely to generate music they found reflective and satisfying.

A total of 8 out of 15 participants modified the song description at least once, typically when the initial version did not align with their expectations. For example, P5 shared, ``\textit{I was expecting something along the lines of joy or relaxation, but when the music was generated, it initially felt lacking in relaxation. The first music, generated based on the initial input, was too fast. However, after modifying it, the tempo slowed down, so I liked it much better.}''

\begin{table}[b]
\caption{Reflectiveness and Satisfaction scores, with and without Tweaking the Song Descriptions}
\Description{A table showing the results of a Mann-Whitney U test comparing Reflectiveness and Satisfaction scores between two groups: Not tweaked and Tweaked. For each metric, the table presents the number of participants, mean, median, standard deviation, and standard error. Tweaked entries show higher mean and median scores than non-tweaked entries in both Reflectiveness and Satisfaction, suggesting a positive effect of tweaking.}
\label{tab:tweak_score}
\centering
\setlength{\tabcolsep}{4pt}
\begin{tabular}{lcccccc}
    \toprule
      & Group         &  N & Mean & Median &    SD &    SE \\
    \midrule
    \multirow{2}{*}{\makecell[l]{Reflectiveness}}
      & Not tweaked   & 35 & 5.26 &   5.25 & 1.169 & 0.198 \\
      & Tweaked       & 15 & 6.05 &   6.50 &  1.150 & 0.296 \\
    \addlinespace
    \multirow{2}{*}{\makecell[l]{Satisfacton}}
      & Not tweaked   & 35 & 5.29 &   6.00 & 0.957 & 0.162 \\
      & Tweaked       & 15 & 6.07 &   7.00 & 1.280 & 0.330 \\
    \bottomrule
  \end{tabular}
\end{table}

Some participants reported that they engaged in exploratory modifications driven by curiosity about how different system settings would affect the musical outcome. P7 explained, ``\textit{To understand the `baseline', I selected `maintain' and listened to the music. After that, I chose both the `moderate' and `amplify' options out of curiosity to see what differences they made.}'' Similarly, others experimented with toggling specific options, such as switching between instrumental and lyrical versions of the same prompt. 

In addition to adjusting settings, a few participants modified the song prompt directly by inserting mood cues, specifying tempo or genre, or adding more detailed instructions to better steer the music generation.. For example, P14 described, ``\textit{I added a note to lower the tempo of the music in song description}.''

However, most participants expressed difficulty in making such modifications due to limited musical knowledge. While they had clear desires to better personalize their music pieces, many felt uncertain about how to express their preferences in musical terms. For example, P2 explained, ``\textit{It was not that I didn’t know how to write a prompt, but I just didn’t have enough knowledge about music genres or musical elements, which made it hard to modify the song description.}'' 

To address these challenges, participants expressed a need for greater system support. Some wanted more structured guidance, such as predefined options or genre suggestions. P4 noted, ``\textit{Even with the same instrumental music, there’s jazz and classical, right? It would be nice if genres like that could be offered as selectable options.}'' Others envisioned more dynamic, context-aware assistance through a conversational interface. P8 described, ``\textit{I wanted more nuanced control, like keeping the overall tone balanced, not too sad or too bright. It felt hard to convey that directly, but a conversational approach, like with Claude or ChatGPT, seemed much easier.}”

\subsubsection{Selection of Emotion-regulation Strategy}

Participants' choice of emotion-regulation strategies showed a significant relationship with the sentiment expressed in their journal entries. A chi-squared test (see Table \ref{tab:sentiment_strategy}) revealed a statistically significant association between journal sentiment and selected emotion-regulation strategy ($\chi^2 = 17.8$, $df = 6$, $p < 0.05$), with distinct patterns emerging across sentiment categories.

For journal entries expressing mixed emotions, the majority (76.9\%) chose to `maintain' their emotional state. When writing about unpleasant emotions, participants most often selected `maintain' (54.5\%), but were also more likely to choose `moderate' option (36.4\%) compared to other sentiment types. In contrast, the `amplify' option was most frequently chosen for pleasant entries, accounting for 52.9\% of those cases, while this was rarely selected outside of positive contexts. For neutral entries, those without explicit emotional content, participants overwhelmingly opted to 'maintain' their neutral emotional state (88.9\%).

\begin{table}[b]
  \caption{Contingency Table of Journal Sentiment and Selected Emotion‑regulation Strategy}
  \Description{A contingency table showing the distribution of emotion-regulation strategies—maintain, moderate, and amplify—across four types of journal sentiment: neutral, pleasant, unpleasant, and mixed. For each sentiment, the observed count and percentage within the row are listed. For example, 88.9\% of neutral entries used the maintain strategy, while 52.9\% of pleasant entries used amplify. The table supports a Chi-squared analysis of association between sentiment and strategy.}
  \label{tab:sentiment_strategy}
  \centering
  \setlength{\tabcolsep}{2pt}
  \begin{tabular}{llrrrr}
    \toprule
    \multicolumn{2}{c}{} 
      & \multicolumn{3}{c}{Emotion‑regulation strategy} 
      & \multicolumn{1}{c}{} \\
    \cmidrule(lr){3-5}
    Sentiment & 
      & maintain & moderate & amplify & Total \\
    \midrule
    Neutral   
      & Observed      &  8  &  1   &  0   &   9  \\
      & \% within row & 88.9\% & 11.1\% &  0.0\% & 100.0\% \\
    Pleasant  
      & Observed      &  6  &  2   &  9   &  17  \\
      & \% within row & 35.3\% & 11.8\% & 52.9\% & 100.0\% \\
    Unpleasant
      & Observed      &  6  &  4   &  1   &  11  \\
      & \% within row & 54.5\% & 36.4\% &  9.1\% & 100.0\% \\
    Mixed     
      & Observed      & 10  &  2   &  1   &  13  \\
      & \% within row & 76.9\% & 15.4\% &  7.7\% & 100.0\% \\
    \midrule
    Total     
      & Observed      & 30  &  9   & 11   &  50  \\
      & \% within row & 60.0\% & 18.0\% & 22.0\% & 100.0\% \\
    \bottomrule
  \end{tabular}
  \Description{Contingency Table of journal sentiment and selected emotion-regulation strategy (Chi-squared analysis)}
\end{table}

Follow-up interviews further illustrated how participants grounded their strategy selections in their emotional states while journaling. For example, P3 shared, ``\textit{I mostly wrote about good things in my journal entry, so I chose to maintain those emotions because I wanted to preserve those moments. However, if I had written something negative, I think I would have selected `moderate' to tone it down}.'' This highlights how participants strategically aligned their choices with their emotional intentions, either to preserve positive feelings or ease negative ones.

\subsubsection{Inclusion of Lyrics}

Participants’ decisions to include lyrics were significantly influenced by the sentiment of their journal entries. A chi-squared test (see Table \ref{tab:sentiment_lyrics}) revealed a statistically significant relationship between journal sentiment and lyrics inclusion  ($\chi^2 = 8.42$, $df = 3$, $p = 0.038$). When journal entries expressed mixed or negative emotions, the majority of participants opted for instrumental music without lyrics (84.6\% and 72.7\%, respectively). In contrast, 64.7\% of entries with positive sentiment were paired with music that included lyrics.

Follow-up interviews shed light on the reasoning behind these choices. Several participants described avoiding lyrics when writing about negative emotions, explaining that lyrics felt too direct or emotionally confrontational. For instance, P5 shared, ``\textit{The lyrics felt like they were directly reflecting what I had written (about my negative emotions), and I didn’t like that as much. I chose instrumental music because I wanted to avoid focusing on specific parts and just let it pass by.}''

Conversely, participants found lyrics helpful for enhancing positive emotional states. P9 explained, ``\textit{The lyrics seemed to enhance the emotions I felt that day while still capturing the essence of those feelings, which I liked}.'' Similarly, P1 described how lyrics elevated the tone of their entry: “\textit{My original emotion was a quiet hope that we could all just live well, maybe at a level of 1—but the music made it feel full of hope, and I really liked that}.” Similarly, P1 highlighted how the lyrics amplified the positivity of her feelings, stating, ``\textit{I’d say it made things more dramatic. My original emotion was just a quiet hope that we could all live well. But when it became music with lyrics, it felt full of hope, which I really liked}.''

\begin{table}[b]
\caption{Contingency Tables of Journal Entry Sentiment and Inclusion of Lyrics}
\label{tab:sentiment_lyrics}
\begin{tabular}{llrrrr}
\toprule
\multicolumn{2}{c}{~} & \multicolumn{2}{c}{Lyrics} & \multicolumn{1}{c}{~} \\
\cmidrule{3-4}
Sentiment & ~             &      Not selected &      Selected &   Total \\
\midrule
Neutral   & Observed      &      5 &      4 &       9 \\
~         & \% within row & 55.6\% & 44.4\% & 100.0\% \\
Pleasant  & Observed      &      6 &     11 &      17 \\
~         & \% within row & 35.3\% & 64.7\% & 100.0\% \\
Unpleasant  & Observed      &      8 &      3 &      11 \\
~         & \% within row & 72.7\% & 27.3\% & 100.0\% \\
Mixed     & Observed      &     11 &      2 &      13 \\
~         & \% within row & 84.6\% & 15.4\% & 100.0\% \\
\midrule

Total     & Observed      &     30 &     20 &      50 \\
~         & \% within row & 60.0\% & 40.0\% & 100.0\% \\
\bottomrule\\
\end{tabular}
\Description{A contingency table presenting the relationship between journal entry sentiment—mixed, negative, positive, and neutral—and the selection of lyrics. For each sentiment category, both the observed count and percentage of entries that selected or did not select lyrics are reported. For example, 64.7\% of positive entries included lyrics, whereas only 15.4\% of mixed entries did. The table supports a Chi-squared analysis to examine the association between sentiment and lyric selection.}

\end{table}

Taken together, these findings illustrate how participants actively shaped their music-augmented journaling experience by engaging with NoRe’s configurable features. From modifying song descriptions to selecting emotion-regulation strategies and lyrics inclusion, users made context-sensitive decisions grounded in their journal content and emotional intentions. These behaviors highlight not only the expressive potential of AI-generated music in reflective practices but also the importance of offering flexible controls that accommodate individual preferences and emotional goals.

\subsection{Perceived Benefits of NoRe}
To evaluate the perceived benefits of using NoRe, we analyzed participants’ responses to three survey items assessing its impact on reminiscence (Q6), emotional regulation (Q7), and self-understanding (Q8). Results were generally positive across all dimensions: participants reported high levels of reminiscence (M = 6.08, SD = 1.07), emotional regulation (M = 5.56, SD = 1.40), and deeper self-understanding (M = 5.36, SD = 1.40).

Follow-up interviews provided further insight into how participants interpreted these benefits. Rather than viewing music generation as a replacement for journaling, participants described NoRe as a meaningful extension of the journaling process. Many viewed the system as a bridge between introspective writing and emotional processing through music. P4 put it: ``\textit{While the journaling process has become a bit longer (with NoRe), it feels worth the added time. Although journaling and meditation are separate activities, NoRe seems to serve as a link between the two}. They viewed NoRe as a medium that connects journal writing with personalized music listening, enhancing the effectiveness of journaling. As such, most participants viewed NoRe as a medium that integrated journal writing with personalized music listening, enhancing the emotional and reflective impact of journaling.

\subsubsection{Stimulating Reminiscence}
Among the three benefit dimensions, reminiscence emerged as the most strongly endorsed. In follow-up interviews, participants frequently described how the generated music enhanced their ability to recall past experiences. For example, P13 reflected: ``\textit{The music from the first day’s journal entry is still vivid in my mind. It feels like the song truly brings back the memory of that day, making it an unforgettable experience.}'' 

Furthermore, participants emphasized that music enabled a more immediate and multisensory connection to past emotions than journaling alone. P6 explained: ``\textit{It’s not just about recording emotions in text; it allows me to quickly revisit how I felt on a specific day. When reflecting on my emotions or events over the past month or week, stimulating multiple senses makes the process much more impactful.}'' Similarly, another participant (P8) noted that music acted as a more intuitive emotional trigger than language: ``\textit{Like visual art, music is more about experiencing and feeling the emotions it evokes before putting them into words. NoRe seems to engage me in these emotional processes, making the experience feel much more immediate.}''


\subsubsection{Enhancing emotional regulation}
Participants also mentioned how listening to AI-generated music after journaling supported emotional regulation. Several noted that music provided an additional step beyond writing, helping them revisit and reassess the emotions they had expressed. For example, P9 remarked: ``\textit{Typically, emotional regulation ends with writing a journal, but listening to the music allows me to reflect on what I wrote and how I felt.}'' Others noted that music facilitated a clearer perspective on their emotional states. P6 shared: ``\textit{Through the music, I was able to recognize how it ultimately impacted my emotional state. Using music as another sensory medium helped me understand and organize my emotions.}''

Beyond reflection, some participants described music as helping them modulate emotions that had intensified during the act of journaling itself. P3 articulated this dynamic: ``\textit{While writing, I tended to recall and ruminate on negative emotions. However, after finishing the journaling and listening to the music, I could experience a process of returning to reality, which helped calm my heightened emotions.}'' This suggests that the AI-generated music might act as a mediator, complementing and balancing the emotional intensity that could arise from deep reflection during journal writing.

In the same vein, the AI-generated music was often perceived as a source of comfort, particularly in response to anxieties or worries expressed in the journal entries. P15 reflected on this, stating, ``\textit{Hearing comforting words related to the concerns I wrote about, accompanied by the music, made me feel really relieved.}'' Some felt the experience went beyond comfort, offering a sense of empathy and understanding. They reported feeling as though the music was responding to their concerns and offering some form of emotional support. For example, P2 shared, ``\textit{It felt like the music aligned well with my emotions. In those cases, it was enjoyable to listen to, and it gave me the sense that someone was empathizing with my feelings, which I really appreciated.}''

Conversely, a few participants noted that the AI-generated nature of the music made it harder to emotionally connect. Specifically, some felt that because the music lacked a human creator’s intention or emotional input, it did not carry the same sense of sincerity or authenticity. P5 put it, ``\textit{The music felt artificial},'' indicating that this perceived artificiality diminished the music’s emotional resonance.

\subsubsection{Deeper Self-Understanding}
Listening to personalized music, knowing it was generated from their own journal entries, led many participants to experience a deeper sense of introspection and self-awareness. For example, P8 reflected, ``\textit{It seems to help me reflect on, understand, and examine my emotions and myself more deeply.}''
Particularly, some participants mentioned that music helped them uncover aspects of their emotional state that they had not consciously recognized while journaling. For instance, P5 shared, ``\textit{It created something I hadn’t thought of myself. [. . .] Listening to the journal-based AI-generated music made me realize and rediscover things I hadn’t noticed before.}’’

Moreover, some participants frequently described the generated music as a mirror-like medium for self-reflection. P11 described this sentiment: ``\textit{What is generated here (in NoRe) is 100\% me.}'' Similarly, P8 remarked, ``\textit{I feel that music that reflects me is more meaningful. As a mirror, I believe that once we express something, we should look back at it. So, it also carries the significance of being a mirror.}'' 

Some even described the music as a more exclusive and emotionally resonant form of self-expression than the journal itself. They emphasized the uniqueness of the music, created from their own writing, as central to its personal meaning. P14 noted, ``\textit{Psychologically, I felt a stronger sense that this was `me' and `mine'.}'' This sense of personal ownership made the music especially impactful in emotionally charged contexts. P9, for instance, reflected on the difficulty of finding existing songs that matched their mood: ``\textit{When experiencing negative emotions, it’s hard to find a pre-existing song that truly resonates. But since this was generated based on my journal, the overall mood aligned more closely with how I felt.}''

Together, these insights suggest that NoRe helped participants deepen their understanding of themselves, not only by reflecting their emotions back to them, but by doing so through a personalized and meaningful medium. This highlights how NoRe could extend the reflective value of journaling by introducing music as an additional way to engage with one’s thoughts and emotions.

\section{Discussion}

In this section, we synthesize findings and reflect on the potential of integrating GenAI for music creation into journaling practices.  We first examine how journal-based music generation may support emotional expression and reflection, highlighting its perceived benefits and usability. We then discuss the dynamic balance between user agency and AI autonomy in the creative process. Finally, we propose design implications and broader ethical considerations for augmenting journaling practices with GenAI.

\subsection{Augmenting Journal Writing with Generative AI for Music Creation}
The in-the-wild evaluation of NoRe offered insights into how AI-generated music could be integrated into everyday journaling practices. Most participants were able to incorporate NoRe into their routines without disrupting their existing habits, suggesting that music augmentation could be layered onto familiar workflows without imposing significant friction. High usability ratings and generally positive feedback further indicate that generative music could function as a viable complement to traditional journaling.

Beyond its potential adoptability, participants also reported that NoRe provided meaningful emotional and reflective benefits.
In particular, three recurring themes emerged: enhanced reminiscence, support for emotional regulation, and deeper self-understanding. Many participants felt that the generated music helped them revisit and reaffirm the emotions and events they had written about. Some also described the music as playing a mediating role—helping to ease heightened emotional states that arose during reflection. These observations resonate with prior research on music's role in emotion regulation and therapeutic processing \cite{enge2022musical,wang2018study,reybrouck2021music}.

Further, some participants described the music as a unique and personal artifact, one that felt distinct from, but meaningfully connected to, their written entries. For many, this sense of personal ownership deepened the music’s reflective value. By pairing introspective writing with personalized music, NoRe appeared to create a synergistic experience that extended the emotional and cognitive benefits of journaling.

Taken together, these findings suggest that journal-based music generation holds promise as an approach for augmenting self-reflection. By bridging written self-expression and auditory emotional engagement, NoRe illustrates how GenAI could support personal introspection and emotional well-being.

\subsection{User Agency and AI Autonomy}
Our findings from both the formative and in-the-wild studies underscore the importance of balancing user agency and AI autonomy in journal-based music generation. In the formative study, participants rated the autonomous prompt highest in perceived reflectiveness, appreciating the AI’s capacity to creatively interpret journal entries with minimal instruction. At the same time, some participants expressed a desire for more influence over musical features, especially to support emotional regulation goals, though notably, none wanted to craft prompts from scratch.

In the in-the-wild evaluation, this preference for a guided-yet-autonomous approach persisted. While a subset of participants engaged in prompt modifications, many reported uncertainty due to limited musical knowledge, reinforcing the need for structured, intuitive controls rather than open-ended customization. Participants who did tweak prompts generally reported higher satisfaction and stronger alignment with their emotional intent, suggesting that lightweight, user-friendly interventions, rather than full control, may offer the best balance. These insights align with prior work on co-creative systems that emphasize scaffolded agency as key to meaningful engagement with GenAI \cite{oh_cocreation, moruzzi_cocreativity, larsson2022towards, inkpen_human, alan_tariff}. 

Based on these findings, we suggest that balancing AI autonomy with structured user support is essential for maximizing satisfaction in journal-based music generation. Future research could explore how varying levels of autonomy affect user satisfaction and emotional resonance. For instance, guided modification features, such as tag-based selection of musical elements or conversational interactions with an AI assistant, may offer intuitive ways for users to influence outcomes. Such flexibility would better accommodate a range of engagement styles, from users who prefer automatic generation to those seeking more hands-on customization.

\subsection{Design Implications of Augmenting Journaling with Generative AI for Music Creation}

Based on findings from the development and evaluation of NoRe, we propose the following design considerations for systems that aim to augment journaling practices through generative AI–based music creation.

\subsubsection{Supporting Users' Existing Journaling Practices}
The design of the journaling interface should aim to support users' existing journaling habits as closely as possible \cite{kim_diarymate_2024}. This includes providing options for adding photos and drawings, which are common elements in traditional journaling practices. To accommodate users who prefer handwritten journals, implementing Optical Character Recognition (OCR) technology could be considered \cite{Hamad2016ADA}, allowing the system to process handwritten entries. This approach would enable both digital and handwritten journal writers to benefit from the music generation feature while maintaining their preferred journaling method.

\subsubsection{Supporting Emotion Regulation}
A central function of journal-based music generation systems like NoRe is to assist users in regulating their emotional states. Participants in our study engaged with journaling for diverse emotional goals—some aimed to maintain their current mood, others to alleviate unpleasant feelings, and some to amplify positive ones. Systems should therefore offer users clear options for specifying their intended emotional direction (e.g., \textit{maintain}, \textit{amplify}, or \textit{moderate}), which can guide the AI's interpretation and generation of music accordingly.

\subsubsection{Balancing AI Autonomy with Guided User Customization}
Rather than expecting users to write detailed prompts, systems should offer structured yet flexible customization options, such as sliders for mood, genre tags, or conversational assistants. These guided tools could make the system accessible to users with limited musical knowledge while still enabling meaningful user control. Prior work highlights how scaffolded co-creation enhances engagement and agency \cite{oh_cocreation, moruzzi_cocreativity, larsson2022towards}. Personalization features that learn and reflect user preferences over time could further streamline the modification experience.

\subsubsection{Supporting Iterative Listening and Reflection}
Participants who compared different music versions or experimented with emotional strategies reported higher satisfaction. To support such exploration, systems should allow users to generate multiple versions, compare them easily, and annotate their reactions. Lightweight refinement tools (e.g., ``too slow,'' ``more energetic'') could help users express preferences without requiring technical knowledge.

\subsubsection{Facilitating Emotional Recollection through Integrated Journal-Music Archives}
Participants found that revisiting journal-linked music deepened emotional recall. To support this, systems should integrate music playback with archived journal entries, enabling users to reinforce multisensory memory and reflection. This multisensory integration could enrich the long-term benefits of journaling by making reflection more vivid and emotionally resonant.

\subsection{Broader Ethical Considerations for Journal-Based AI Music Systems}
While our findings demonstrated the potential of augmenting journaling with AI-generated music, it also surfaced broader questions about the responsible deployment of GenAI in emotionally sensitive domains. In this section, we outline three key ethical considerations that may inform the future development and application of journal-based AI music systems, particularly as they scale or move into clinical and high-stakes settings.

\subsubsection{Emotional Risks in Long-Term or Clinical Use}
Although participants found NoRe beneficial for emotional reflection and regulation, sustained self-reflection, especially without human oversight, could sometimes exacerbate distress, pessimism, or over-introspection \cite{lengelle_is_2016, woerkom_critical_2010}. These risks could be amplified when such systems are used by vulnerable individuals or considered for clinical contexts. To ensure safe deployment, future adaptations should incorporate safeguards such as clear disclaimers, escalation protocols, and integration with human support systems, as recommended in prior work \cite{meadi_exploring_2025, de_freitas_chatbots_2024, babu_digital_2025}.



\subsubsection{Privacy and Data Governance}
Journal-based AI systems like NoRe process highly sensitive and intimate user content, making privacy a central ethical concern. During our study, participants often disclosed deeply personal reflections, highlighting the importance of secure data handling. When such systems rely on general-purpose cloud-based LLMs or third-party APIs, the risk of unintended data exposure increases \cite{kim_mindfuldiary_2024}. Deploying local LLMs offers a promising alternative by keeping user data on personal or institutional hardware, thus reducing external exposure \cite{perron_moving_2024, wiest_privacy-preserving_2024}. However, this approach also introduces practical challenges, including the need for greater infrastructure investment and ongoing maintenance.

Beyond technical approaches, fostering trust requires transparent communication about data practices \cite{Zhang2024, varghese2024public}. Users should be clearly informed about how their data is collected, stored, and used—and be given control over these processes. Options to opt out of data storage, delete past entries, or set boundaries on content processing could empower users and promote a greater sense of psychological safety.

\subsubsection{Ethical Concerns in AI-generated Music Ecosystems}
Because NoRe leverages AI-generated music, it intersects with broader debates surrounding copyright, creative attribution, and the displacement of artistic labor \cite{drott_copyright_2021, militsyna_human_2023}. While the system is designed for personal and reflective use, its reliance on models trained on large musical corpora raises questions about how creative works are sourced and whether artists are fairly compensated \cite{jacques_protecting_2024}. Therefore, even in therapeutic or non-commercial contexts, designers must remain accountable for the systemic implications of their tools. Responsible integration of generative music systems thus requires attention not only to user benefit but also to the creative ecosystems these technologies impact.


\subsection{Limitations and Future Work}
Our study has several limitations. First, all our participants were Korean-speaking undergraduate or graduate students between the ages of 19 and 29, and were predominantly female. This could raise concerns regarding sampling bias and generalizability. Future research should address these demographic limitations by recruiting participants from diverse backgrounds and age groups. 

Second, we employed a single-arm study design without a control group. As such, all statistical findings are descriptive and correlational, and we cannot draw causal conclusions about NoRe’s effectiveness compared to alternative journaling or music-based interventions. Future research should incorporate controlled experimental designs to better evaluate comparative efficacy.

Third, the study spanned only seven days, which may be insufficient to assess the long-term effects of journal-based AI-generated music. Longitudinal studies are needed to examine sustained emotional or reflective benefits over time.

Finally, we used OpenAI’s GPT-4o and Suno for music generation. As GenAI models evolve rapidly, results may vary with future model updates. Future studies should explore the robustness of journal-based music systems across different model architectures and versions to assess generalizability and reproducibility.


\section{Conclusion}
This paper introduced NoRe, a system that augments journaling with AI-generated music to support emotional reflection, regulation, and self-understanding. Drawing on formative findings about the emotional and archival value of journaling, we designed NoRe to translate users’ written entries into personalized musical responses. Findings from a formative and a seven-day in-the-wild study showed that participants appreciated the emotional relevance of the generated music and often described it as a meaningful extension of their journal writing. Many described the AI-generated music as deeply personal, which enhanced their emotional engagement and prompted new forms of introspection.

Building on these insights, we proposed several design considerations for future systems that integrate generative AI into reflective practices: supporting users’ existing journaling habits, assisting emotion regulation, providing guided customization that balances user agency with AI autonomy, enabling iterative listening and refinement, and facilitating emotional recollection through integrated journal–music archives.

Taken together, our findings highlight the potential for generative AI to expand how people engage with personal reflection. This work contributes empirical insights and design guidance for future systems that aim to support introspection, emotional processing, and self-understanding through collaborative human-AI experiences.

\begin{acks}
We thank our study participants and the reviewers for their valuable feedback and contributions. We also extend our gratitude to Jiin Cheon for his earlier involvement in this project. This work was supported by the Student-Directed Education Program, the New Faculty Startup Fund from Seoul National University (Grant No. 200-20230022), and the Undergraduate Research Learner (URL) Program of Information Science and Culture Studies at Seoul National University.
\end{acks}

\bibliographystyle{ACM-Reference-Format}
\bibliography{base}

\appendix
\clearpage
\appendix

\onecolumn

\section{Formative Participants}
\label{appendix:formativeparticipant}

\begin{table}[H]
  \centering
  \caption{Summary of our formative study participants. The (ID) column indicates the ID assigned to each participant. The (Age) column denotes their age, and the (Gender) column indicates gender (M: male; F: female). The Journaling Duration column shows how long each participant has been engaged in journaling. The Frequency of Journaling column reflects how often they write—Daily (almost every day), Regularly (2–3 times per week), or Occasionally (about once a month). The final two columns indicate whether participants had prior experience listening to music generated by AI or creating music using AI. ‘O’ indicates yes; ‘X’ indicates no.}
  \Description{A table presenting the summary of 15 participants from the formative study. The (ID) indicates the ID assigned to participants. The (Age) column denotes the age of the participants. The (Gender) column shows the gender of participants (M: male; F: female). The Journaling Duration column indicates how long each participant has been journaling. The Frequency of Journaling column reflects how often participants wrote in their journals—Daily (almost every day), Regularly (2–3 times per week), or Occasionally (about once a month). The final two columns indicate whether participants had experience listening to AI-generated music or creating music using AI. ‘O’ indicates yes; ‘X’ indicates no.}
  \begin{tabular}{cccccccc}
    \toprule
    \textbf{ID} & \textbf{Age} & \textbf{Gender} & \textbf{Journaling Duration}
      & \textbf{Frequency of Journaling} &  \makecell{\textbf{AI Music}\\\textbf{Listening Experience}} & \makecell{\textbf{AI Music}\\\textbf{Generation Experience}} \\
    \midrule
    F1  & 25 & Female & 2 years    & Occasionally & O & X \\
    F2  & 26 & Male   & 8 years    & Regularly    & X & X \\
    F3  & 25 & Male   & 4 years    & Regularly    & O & X \\
    F4  & 25 & Female & 16 years   & Daily        & X & X \\
    F5  & 24 & Female & 6 months   & Daily        & O & X \\
    F6  & 26 & Female & 10 years   & Regularly    & X & X \\
    F7  & 25 & Female & 4 years    & Regularly    & X & X \\
    F8  & 26 & Female & 6 years    & Regularly    & O & O \\
    F9  & 22 & Female & 3 years    & Occasionally & X & X \\
    F10 & 20 & Female & 3 years    & Occasionally & O & O \\
    F11 & 23 & Female & 2 years    & Daily        & O & X \\
    F12 & 24 & Female & 6 years    & Daily        & X & X \\
    F13 & 22 & Female & 6 years    & Occasionally & O & O \\
    F14 & 25 & Female & 3 years    & Daily        & O & X \\
    F15 & 22 & Male   & 8 months   & Daily        & O & O \\
    \bottomrule
  \end{tabular}
  \label{tab:formative-participants}
\end{table}

\section{In-the-Wild Evaluation Participants}
\label{appendix:evaluationparticipant}

\begin{table}[H]
  \caption{Summary of our in-the-wild evaluation study participants. The (ID) column indicates the ID assigned to each participant. The (Age) column denotes their age, and the (Gender) column shows their gender (M: male; F: female). The Journaling Duration column indicates how long each participant has been engaged in journaling. The Frequency of Journaling column reflects how often they typically wrote—Daily (almost every day), Regularly (2–3 times per week), or Occasionally (about once a month). The final two columns indicate whether participants had prior experience using large language models (LLMs, such as ChatGPT) or creating music with AI tools. ‘O’ indicates yes; ‘X’ indicates no.}
  \Description{A table presenting the summary of 15 participations form the in-the-wild evaluation study. The (ID) indicates the ID assigned to participants. The (Age) column denotes the age of the participants. The (Gender) column shows the gender of participants (M: male; F: female). The Journaling Duration column indicates how long each participant has been journaling. The Frequency of Journaling column reflects how often participants wrote in their journals—Daily (almost every day), Regularly (2–3 times per week), or Occasionally (about once a month). The final two columns indicate whether participants had experience using large language models (LLMs), or had created music using AI. ‘O’ indicates yes; ‘X’ indicates no.}
  \begin{tabular}{cccccccc}
    \toprule
    \textbf{ID}
      & \textbf{Age}
      & \textbf{Gender}
      & \textbf{Frequency of Journaling}
      & \makecell{\textbf{AI Proficiency}\\\textbf{(7-point Likert Scale)}}
      & \makecell{\textbf{LLM}\\\textbf{Experience}}
      & \makecell{\textbf{AI Music}\\\textbf{Generation Experience}} \\
    \midrule
    P1  & 25 & Female & Daily & 3 & O & X \\
    P2  & 25 & Female & Occasionally & 4 & O & X \\
    P3  & 25 & Female & Occasionally & 7 & O & O \\
    P4  & 26 & Female & Occasionally & 3 & O & X \\
    P5  & 22 & Female & Occasionally & 5 & O & X \\
    P6  & 21 & Male & Regularly & 4 & O & X \\
    P7  & 19 & Male & Occasionally & 6 & O & X \\
    P8  & 26 & Female & Daily & 5 & O & X \\
    P9  & 20 & Female & Occasionally & 5 & O & O \\
    P10 & 24 & Female & Occasionally & 5 & O & X \\
    P11 & 28 & Female & Regularly & 5 & O & X \\
    P12 & 21 & Female & Occasionally & 4 & O & O \\
    P13 & 26 & Female & Regularly & 5 & O & X \\
    P14 & 24 & Female & Daily & 4 & O & X \\
    P15 & 29 & Female & Regularly & 4 & O & X \\
    \bottomrule
  \end{tabular}
  \label{tab:evaluation-participants}
\end{table}

\clearpage

\end{document}